\documentstyle[aps,epsfig]{revtex}
\begin{document}
\twocolumn[\hsize\textwidth\columnwidth\hsize\csname@twocolumnfalse\endcsname
\draft
\title{From avalanches to fluid flow: A continuous picture of grain
dynamics down a heap}
\author{P.-A. Lemieux and D. J. Durian}
\address{UCLA Department of Physics \& Astronomy, Los Angeles, CA 90095-1547}
\date{\today}
\maketitle

\begin{abstract}
    Surface flows are excited by steadily adding spherical glass beads
    to the top of a heap.  To simultaneously characterize the fast
    single-grain dynamics and the much slower collective intermittency
    of the flow, we extend photon-correlation spectroscopy via
    fourth-order temporal correlations in the scattered light
    intensity.  We find that microscopic grain dynamics during an
    avalanche are similar to those in the continuous flow just above
    the transition.  We also find that there is a minimum jamming
    time, even arbitrarily close to the transition.
\end{abstract}

 \vskip2pc]
 \narrowtext

We have all poured sand in a heap.  If grains are added rapidly, the
heap relaxes through continuous, fluid-like flow at its surface.  If
grains are added slowly, it relaxes intermittently through discrete
avalanches.  The existence of both continuous and intermittent regimes
is characteristic of the dual nature of granular media, which can
exhibit properties of both liquids and solids \cite{jaeger96,duran99}.
Similar behavior, such as bubble rearrangements in foams or kinetic
heterogeneities in supercooled liquids, arises in other systems near
the onset of jamming \cite{cates98,liu98}.  In spite of its ubiquity,
and its obvious importance for applications and geophysical phenomena,
the transition between intermittent and continuous flow has received
relatively little attention.  One reason may be that kinetic theories
\cite{jenkins83,haff83} of granular hydrodynamics, as well as
minimal-ingredient theories of surface behavior \cite{bouchaud94},
apply only to continuous flows, while cellular-automata models of
avalanches apply only to rate-independent intermittent flows
\cite{bak88}; both approaches break down near the transition.  Another
reason may be the vast separation in time scales between the grain
velocity fluctuations and the intermittency dynamics, which makes it
difficult to capture both phenomena simultaneously.

Recently we used diffusing-wave spectroscopy (DWS), a
photon-correlation spectroscopy for opaque media \cite{dws}, to probe
motion in homogeneously excited sand \cite{menon97ab}.  DWS reveals
dynamics on time- and length-scales too small for imaging, but cannot
be applied to intermittent flows.  This is because in the static
periods all the grains are perfectly correlated, violating the
assumption that the total scattered electric field has Gaussian
statistics.  Here, for a heap upon which grains are steadily poured,
we show how intermittency can be detected and characterized using
higher-order temporal correlations in the intensity of
multiply-scattered laser light.  We extract single-grain dynamics, as
well as probability functions describing the collective intermittency
of the flow.  This gives an unprecedented picture of the dynamics in
both intermittent and continuous flows, as well as across the
transition between these extremes.

\begin{figure}
\centerline{\epsfig{file=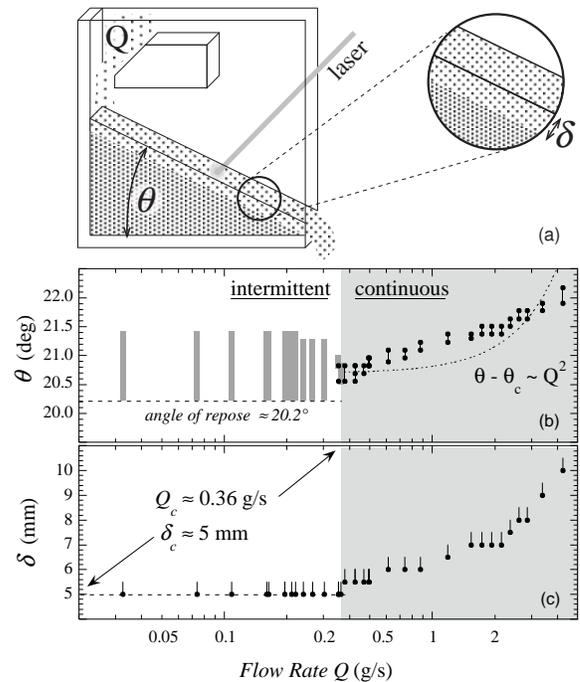,width=3.00in}}
\caption{The experimental setup (a), and two macroscopic measures of
the response vs mass flow rate: (b) heap angle and (c)
layer thickness.}
\label{setup}
\end{figure}

\paragraph*{System.} The experimental setup (Fig.~\ref{setup}a) consists of
parallel static-dissipating plastic walls, 30x30~${\rm cm}^{2}$,
forming a $w=9.5$~mm wide vertical channel closed at the bottom and on
one side.  We use spherical glass beads of diameter $0.33\pm 0.03$~mm,
large enough to minimize effects of interstitial gas and electrostatic
forces while keeping the system size manageable; the density of the
packing is $\rho=1.35$~g/cc.  Flow is created by dropping grains into
the closed side of the channel, and breaking their fall by a secondary
hopper just above the heap.  Grains falling out the open end are
recycled by a vacuum-driven elevator.  The flow rate is set by a
micrometer-controlled gate valve and a knife-edge, allowing for wide
variation of 0.025-5~g/s (50-10,000 grains/s).  Prior to a run, the
beads are washed, dried in air, and circulated until a steady state is
reached.

For orientation the macroscopic behavior of the heap is summarized in
Figs.~\ref{setup}b-c vs mass flow rate, $Q$, the sole parameter that
sets the grain dynamics.  As opposed to the inclined-plane geometry
\cite{pouliquen99}, our system chooses both the angle $\theta$ of the
heap and the thickness $\delta$ of the flowing layer.  This avoids
such issues as the role of the plane roughness, and should ultimately
permit more direct contact with theory.  Visual inspection reveals
that the transition between intermittent and continuous flows is at
$Q_{c}=0.36$~g/s, and that it is rather sharp.  By contrast with the
rotating-drum geometry \cite{jaeger89,rajchenback90}, this transition
is not hysteretic \cite{rajchenback90}, and avalanches always start at
the top.  Also by contrast, both $\delta$ and $\theta$ are
uniform across the whole surface.  Fig.~\ref{setup} shows that these
quantities become independent of $Q$ in the intermittent regime, with
$\theta$ ranging between a minimum angle of repose and a maximum angle
of stability, and with $\delta$ being the same thickness as at the
transition, $\delta_{c}=5$~mm, which is about 15 grain diameters or
about half the width of the channel.  Thinner layers cannot flow
indefinitely, as dissipative forces overwhelm gravity and inertia.
Fig.~\ref{setup} also shows how both $\theta$ and $\delta$ increase
monotonically for continuous flows above the transition.
Interestingly, the heap angle varies logarithmically with $Q$ and, as
measured at the walls for $Q=0.5$~g/s, the velocity vs depth is nearly
exponential, $v_{x}(z)\approx(1~{\rm cm/s})\exp[-z/(0.15~{\rm cm})]$
for $z<\delta$ \cite{profile}.  This is unlike the modified-Bagnold
model predictions of $\theta-\theta_{c}\propto Q^{2}$ and $v_{x}(z)
\propto (\delta^{3/2}-z^{3/2})$ \cite{rajchenback90}.  This implies
that there must be enduring contacts and/or that the rate of
collisions is not set by ${\rm d}v_{x}/{\rm d}z$.

\paragraph*{Light Scattering.} While visual measurements provide a rough
overview of a heap's macroscopic response, multiple-light scattering
measurements provide a detailed picture of both individual and
collective grain dynamics over a broad range of time scales.  Here we
illuminate the top surface with an ${\rm Ar}^{+}$ laser, wavelength
$\lambda=514$~nm, over a 8~mm region located $L=18$~cm from the closed
end.  The intensity fluctuations in the speckle pattern created by
backscattered photons are captured by a single-mode optical fiber
coupled to a photon-counting assembly.  Following
Ref.~\cite{lemieux99}, we measure simultaneously in real-time both the
second- and fourth-order intensity correlation functions
$g^{(2)}(\tau)=\left< I(0)I(\tau) \right> / \left< I \right>^{2}$ and
$g^{(4)}_{T}(\tau)=\left< I(0)I(T)I(\tau)I(\tau+T) \right> / \left< I
\right>^{4}$, where $T$ is a fixed delay that can be chosen between
50~ns and 50~ms, and $\tau$ is swept over a range of values by a
digital correlator.  In analogy with the `second-spectrum'
\cite{weissman96}, $g^{(4)}_{T}(\tau)$ can be thought of as a
`second-correlation' or `correlation of correlation' quantifying
fluctuations in the correlation signal $I(t)I(t+T)$ with time $t$.

\begin{figure}
\centerline{\epsfig{file=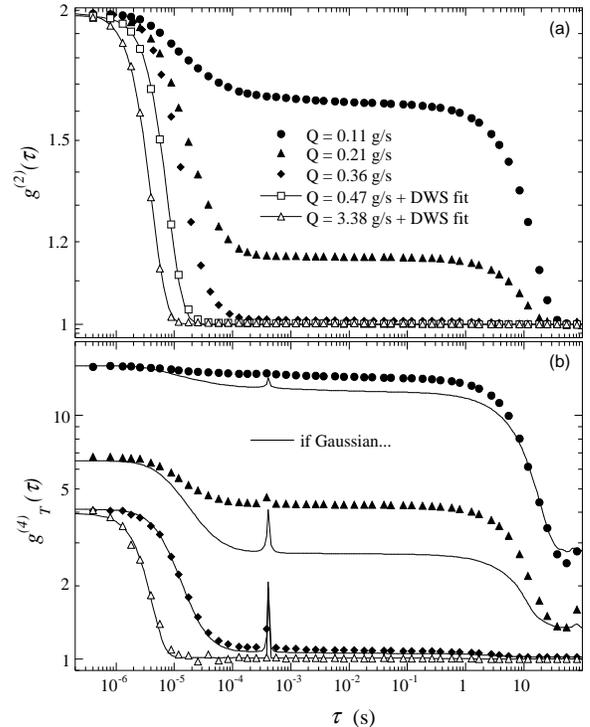,width=3.00in}}
\caption{Examples of (a) second- and (b) fourth-order temporal
correlations in the intensity of laser light multiply-scattered from
the surface of the heap.  For continuous flows, the fluctuations are
Gaussian and the correlations exhibit a single decay; for intermittent
flows, the fluctuations are non-Gaussian and the correlations exhibit
a two-step decay.}
\label{light}
\end{figure}

Examples of $g^{(2)}(\tau)$ and $g^{(4)}_{T}(\tau)$ are shown in
Fig.\ref{light}.  For continuous flows, both correlations exhibit a
single complete decay at short time scales.  Good agreement is seen
between $g^{(4)}_{T}(\tau)$ data and the predictions generated solely
from $g^{(2)}(\tau)$ data, as in Ref.~\cite{lemieux99}, showing that
the fluctuations are Gaussian.  Therefore, the usual machinery of DWS
may be applied to extract single-grain dynamics.  For intermittent
flows, by contrast, both correlations exhibit a partial decay at short
times, followed by a plateau and a final decay at long times.
Furthermore, the fluctuations are non-Gaussian, implying
that a new model is required to extract grain motion.

Several ingredients are required to model the full spectrum of grain
dynamics.  One is the usual normalized electric field autocorrelation
function $\gamma(\tau)$, whose Fourier transform gives the broadening
of the power spectrum due to relative motion of scattering sites
during flow.  For our backscattering geometry, the theory of DWS
\cite{dws,wu90} gives $|\gamma(\tau)|\propto\exp[-2\sqrt{k^{2}\delta
v^{2}\tau^{2}+3l^{*}/l_{a}}]$, where $k=2\pi/\lambda$, $\delta v$ is a
typical relative grain speed averaged over the scattering volume,
$l^{*}=1.3$~mm is the photon transport mean-free path, and
$l_{a}=1.0$~cm is the photon absorption length.  Since $\delta v$ is
the only quantity not known in advance, it may be extracted from the
data.

All other ingredients relate to the probabilities with which the
system switches between flowing / `1' and static / `0' states.  Most
important is the probability $P_{0}(\tau)$ for the system to be in the
{\it same} static configuration after a time interval $\tau$.  This function
decays monotonically from one to zero.  It can be related to the
probability $\tilde P(\tau)$ that two flowing states are separated by
$\tau$ via $P_{0}(\tau) = \int^{\infty}_{\tau}\tilde P(\tau')
(\tau'-\tau) {\rm d}\tau'/t_{0}$, and hence by $\tilde P(\tau)=t_{0}{\rm
d}^{2}P_{0}/{\rm d}\tau^{2}$, where $t_{0}$ is the average static
state duration.  Also important are the probabilities $P_{ij}(\tau)$
for the system, if initially in state $i$, to be in state $j$ after a
time $\tau$.  These interrelated functions decay from either 1 or 0 to
the fraction of time $f_{j}$ spent in state $j$.

Making the simplest assumption consistent with visual observations,
that starting/stopping transients are neglibible and that the speckle
pattern fluctuates rapidly during each flowing state but is constant during
each static state, we find
\begin{equation}
    g^{(2)}(\tau) = 1 + \beta \left[ f_{1}P_{11}(\tau)|\gamma(\tau)|^{2}
    + f_{0}P_{0}(\tau) \right],
\label{g2}
\end{equation}
where $1/\beta$ is roughly the number of speckles viewed by the
detector.  For continuous flow, where $f_{1}$ and $P_{11}$ are
identically one, Eq.(\ref{g2}) reduces to the usual Siegert relation
connecting intensity and field autocorrelations.  For intermittent
flow, Eq.(\ref{g2}) exhibits a decay at short times due to
$\gamma(\tau)$, during which the switching functions are constant, and
a final decay at long times due to $P_{0}(\tau)$.  This is unlike the
two-step decay seen in supercooled systems, where the field
fluctuations are Gaussian and the final decay (the
$\alpha$-relaxation) is due to cage-hopping effects.  Note that the
intercept $g^{(2)}(0)=1+\beta$ gives the same second intensity moment
as for a Gaussian process; this property also holds in our general
expressions for $g^{(3)}$ and $g^{(4)}$.  Intermittency gives
non-Gaussian fluctuations because of {\it grain correlations} during
each static state, unlike the broader intensity distributions familiar
from number fluctuations in single-scattering experiments.

The predictions for higher-order correlations are cumbersome, but
reduce considerably for the usual case of good separation in decay
scales for $\gamma(\tau)$ and the switching functions.  Choosing an
intermediate value for the fixed delay $T$, so that $\gamma(T)=0$ and
$P(\tau\pm T)=P(\tau)$, we find
\begin{eqnarray}
    g^{(4)}_{T}(\tau)
    & \approx & 1 + \beta f_{1}\left[ 2 + \beta |\gamma(\tau)|^{2} \right]
                      |\gamma(\tau)|^{2} + \nonumber \\
  \beta & f_{0} &  \left[ 2+\beta P_{00}(t) +
                       \left( 4+10\beta+6\beta^{2}\right) P_{0}(t) \right].
\label{g4}
\end{eqnarray}
This neglects terms in $\gamma(|\tau-T|)$, which cause a peak in
$g^{(4)}_{T}(\tau)$ at $\tau=T$ that is characteristic of our slice of
the full $g^{(4)}(\tau_{1},\tau_{2},\tau_{3})$.  As above,
Eq.(\ref{g4}) exhibits a single decay for continuous flows and a
two-step decay for intermittent flows.  Armed with
Eqs.~(\ref{g2}-\ref{g4}), one can not only reliably detect the
presence of intermittency, but can also characterize both the grain
motion via $\gamma(\tau)$ and the collective intermittency of the
flows via $P_{0}(\tau)$ and $P_{ij}(\tau)$.  This extends the utility
of photon-correlation techniques (not just DWS) to systems like
slowly-driven sand exhibiting intermittency.

\begin{figure}
\centerline{\epsfig{file=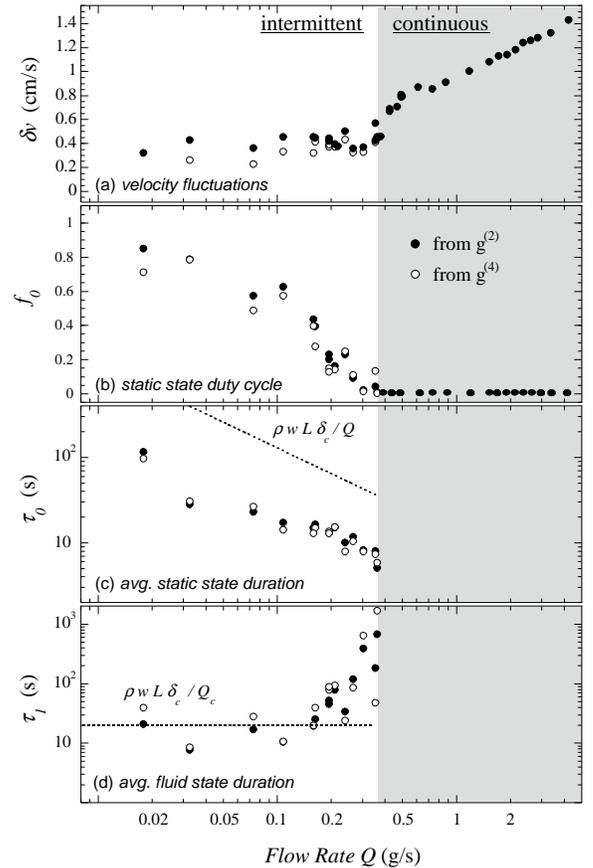,width=3.00in}}
\caption{Grain dynamics vs mass flow rate: (a) velocity fluctuations,
(b) fraction of time spent static, (c) average time spend static, and
(d) average time spend flowing.  Consistent results are obtained from
second- and fourth-order correlation data using Eqs.\ref{g2} and
\ref{g4}, respectively.}
\label{sand}
\end{figure}

\paragraph*{Grain Dynamics.}  For continuous flows, according to the
above discussion, the correlations fully decay at a time scale given
approximately by $\lambda/\delta v$.  Using $\delta v$ as the sole
parameter, we obtain excellent fits to $g^{(2)}(\tau)$ (and
equivalently to $g^{(4)}_{T}(\tau)$ since here the fluctuations are
Gaussian) as shown in Fig.~\ref{light}(a).  The results in
Fig.~\ref{sand}(a) show that $\delta v$ varies logarithmically with
the flow rate, just like the heap angle.  The value of $\delta v$
appears to reflect both shear flow and random ballistic motion.  At
$Q=0.5$~g/s, slightly above the transition, the r.m.s. average strain
rate in the scattering volume is estimated from $v_{x}(z)$ and $Q$ to
be less than 4/s.  This gives a contribution to $\delta v$ of
$l^{*}\sqrt{\left< \dot{\gamma}^2 \right> /5}=0.4$~cm/s \cite{wu90},
which accounts for half the fitted value of 0.8~cm/s.  We believe that
the random contribution is not due to free-flight between grain-grain
collisions, since the Bagnold model does not apply, but is rather due
to jostling parallel and perpendicular to the average flow as shearing
layers slide over one another with grains always maintaining a near
close-packing arrangement.  While there is no saltation, altered
dynamics at the free surface may also contribute.  In any case, the
extracted values of $\delta v$ represent a measure of microscopic
velocity fluctuations for the flowing grains.  These are related to
the average flow speed as $\delta v \sim v^{0.5\pm0.1}$, similar to
previous observations for a vertical hopper \cite{menon97ab}.  The
increase of fluctuations relative to the average may be the ultimate
cause of jamming at low driving rates.

For intermittent flows, according to the light scattering discussion,
the correlations partially decay at $\lambda/\delta v$ and then fully
decay at a time scale reflecting the intermittency dynamics.
Eqs.~(\ref{g2}-\ref{g4}) show that between these decays the
correlations plateau at values of $1+\beta f_{0}$ and $1 + \beta
(6+11\beta+6\beta^{2})f_{0}$, respectively.  Knowing $\beta \approx
1$, we easily extract the fraction of time $f_{0}$ the system spends
at rest.  The results in Fig.~\ref{sand}(b) show that the
$g^{(2)}(\tau)$ and $g^{(4)}_{T}(\tau)$ plateaus are consistent and in
accord with expectations.  As the flow rate goes to 0 ($Q_{c}$),
$f_{0}$ approaches 1 (0), denoting a completely static (fluid) state.

Given the correlation plateaus, we may also fit the early-time decay for
grain-scale velocity fluctuations {\it during} avalanches using
Eqs.(\ref{g2}-\ref{g4}).  The results in Fig.~\ref{sand}(a) show that
$\delta v$ is approximately independent of $Q$ in the intermittent
regime.  Furthermore, the value is close to, but perhaps slightly less
than, that for fluctuations in continuous flow just above the
transition.  So avalanches are not very different from
barely-continuous flow.

Finally we analyze the late-time decay of the intensity correlations
in terms of avalanche statistics.  Specifically, the functional form
of the the switching function $P_{0}(\tau)$ may be deduced from
$g^{(2)}(\tau)$ data using Eq.(\ref{g2}) and the known plateau values.
Similarly $P_{0}(\tau)$ may be deduced from $g^{(4)}_{T}(\tau)$ data,
ignoring the small contribution in Eq.(\ref{g4}) from $P_{00}(\tau)$.
The two results agree, and for all $Q$ are consistent with the
empirical form $P_{0}(\tau) = \exp(-\tau/t_{0}) /
[1+(\tau/t_{0})^{2}/2]$.  Namely, the first half of the decay is
linear, and the second half is faster than exponential.  This rules
out self-organized criticality, for which $P_{0}(\tau)$ would be a
power-law, as well as random-telegraph switching, for which
$P_{0}(\tau)$ would be perfectly exponential.  Rather, the
distribution of intervals between avalanches is peaked around an
average, $t_{0}$, similar to behavior seen in Ref.~\cite{jaeger89}.

Results for the average static and flowing times, deduced respectively
from $P_{0}(\tau) = 1-\tau/t_{0}+\ldots$ and $f_{0} = t_{0} /
(t_{0}+t_{1})$, are shown in Fig.~\ref{sand}(c-d) as a function of
flow rate.  For the smallest $Q$, we find evidence of a quasi-static
limit in which the time between avalanches scales as $1/Q$ and in
which the duration of avalanches $t_{1}$ approaches a constant.  The
observed times are on the order of the upper limits $\rho
wL\delta_{c}/Q$ and $\rho\delta_{c}wL/Q_{c}$ needed to fill and to
empty, respectively, a volume $wL\delta_{c}$ of sand above the
illumination spot.  As the flow rate is increased toward $Q_{c}$,
where continuous flow commences, the time spent flowing appears from
Fig.~\ref{sand}(d) to diverge.  Interestingly, at the same point, the
time spent in the static state does not vanish continuously but rather
approaches a nonzero constant.  Even arbitrarily close to smooth flow,
it takes significant time for the system to start moving again after
it has jammed.  This is analogous to common experience on a freeway,
where traffic jams develop very suddenly but clear much more slowly.

{\it Conclusions.} In summary we have introduced a novel optical
method and used it to probe, simultaneously, both avalanche
statistics and microscopic grain dynamics {\it within} the avalanches.
A unique feature is that the technique applies to both continuous and
quasi-static intermittent flows, as well as to the transition between
these extremes.  This should enable a wide range of new experiments on
granular dynamics as a high driving rate of tumbling, shaking,
blowing, sprinkling, shearing, etc., is decreased and the system
develops intermittent dynamics as a precursor to complete
jamming.  For example we are currently investigating the influence of
channel dimensions, and the behavior as a function of depth, for the
sprinkling geometry presented here.  The results should help guide and
test fundamental theories of granular flow that are based on the
actual micrsocopic grain dynamics.

This work was supported by NSF grant DMR-0070329.

%

%
%

%
%

\end{document}